\documentclass[conference]{IEEEtran}
\usepackage{graphicx}
\usepackage[T1]{fontenc}
\usepackage[]{caption}
\usepackage{amsmath}
\usepackage{array}
\ifCLASSINFOpdf
\else
\fi
\begin{document}
%
\title{To Love or to Loathe: How is the World Reacting to China's Rise?}

\author{\IEEEauthorblockN{Yu Wang}
\IEEEauthorblockA{Department of Political Science\\
University of Rochester\\
Rochester, NY, 14627, USA\\
\textit{ywang176@ur.rochester.edu}}
\and
\IEEEauthorblockN{Jianbo Yuan}
\IEEEauthorblockA{Department of Computer Science\\
University of Rochester\\
Rochester, NY, 14627, USA\\
\textit{jyuan10@ur.rochester.edu}}
\and
\IEEEauthorblockN{Jiebo Luo}
\IEEEauthorblockA{Department of Computer Science\\
University of Rochester\\
Rochester, NY, 14627, USA\\
\textit{jluo@cs.rochester.edu}}
}


%


\maketitle

\begin{abstract}
China has experienced a spectacular economic growth in recent decades. Its economy grew more than 48 times from 1980 to 2013. How are the other countries reacting to China's rise? Do they see it as an economic opportunity or a security threat? In this paper, we answer this question by analyzing online news reports about China published in Australia, France, Germany, Japan, Russia, South Korea, the UK and the US. More specifically, we first analyze the frequency with which China has appeared in news headlines, which is a measure of China's influence in the world. Second, we build a Naive Bayes classifier to study the evolving nature of the news reports, i.e., whether they are economic or political. We then evaluate the friendliness of the news coverage based on sentiment analysis. Empirical results indicate that there has been increasing news coverage of China in all the countries under study. We also find that the emphasis of the reports is generally shifting towards China's economy. Here Japan and South Korea are exceptions: they are reporting more on Chinese politics. In terms of global sentiment, the picture is quite gloomy. With the exception of Australia and, to some extent, France, all the other countries under examination are becoming less positive towards China.
\end{abstract}


%
\IEEEpeerreviewmaketitle

\section{Introduction}
China's rise promises to be one of the great dramas of the 21st century. Its economy grew more than 48 times from 1980 to 2013. Figure 1 shows the dramatic growth of the Chinese GDP from 1961 to 2013 relative to other major economies.\footnote{GDP figures are from the World Bank, available at http://data.worldbank.org/indicator/NY.GDP.MKTP.CD.} Economists argue about whether China will become the world's largest economy; political scientists argue about whether China's rise can be peaceful and beneficial for the rest of the world. It is of great importance to understand how countries around the world are reacting to the increasingly prosperous and assertive China: do they perceive China's rise as an opportunity or a threat?

In this paper, we focus on eight representative countries: Australia, France, Germany, Japan, South Korea, Russia, the UK and the US. The US is the world's only current superpower. France, Germany and the UK are Europe's three largest economies. Japan and South Korea are China's close neighbors and China is their largest trading partner. Russia is a former superpower and has a strong economic, diplomatic, and military relationship with China. Australia is culturally Western but geographically in the East. China is Australia's largest trading partner. 

Gauging public opinion is usually based on public opinion polls. But time series poll data about attitudes towards China is scarce. Opinion polls regarding China rarely stretch back more than ten years. Examining newspapers offers an alternative: the frequency with which China appears in news headlines can serve as a measure of China's global influence; classifying news articles on China as either economic or political can shed light on the evolving nature of foreign perspectives on China; and the China Friendliness Index (proposed in this work), based on sentiment analysis, can reveal the evolution of China's bilateral relations with the countries of interest.

\begin{figure}[h!]
\centering
\includegraphics[width=8.9cm, height=6.5cm]{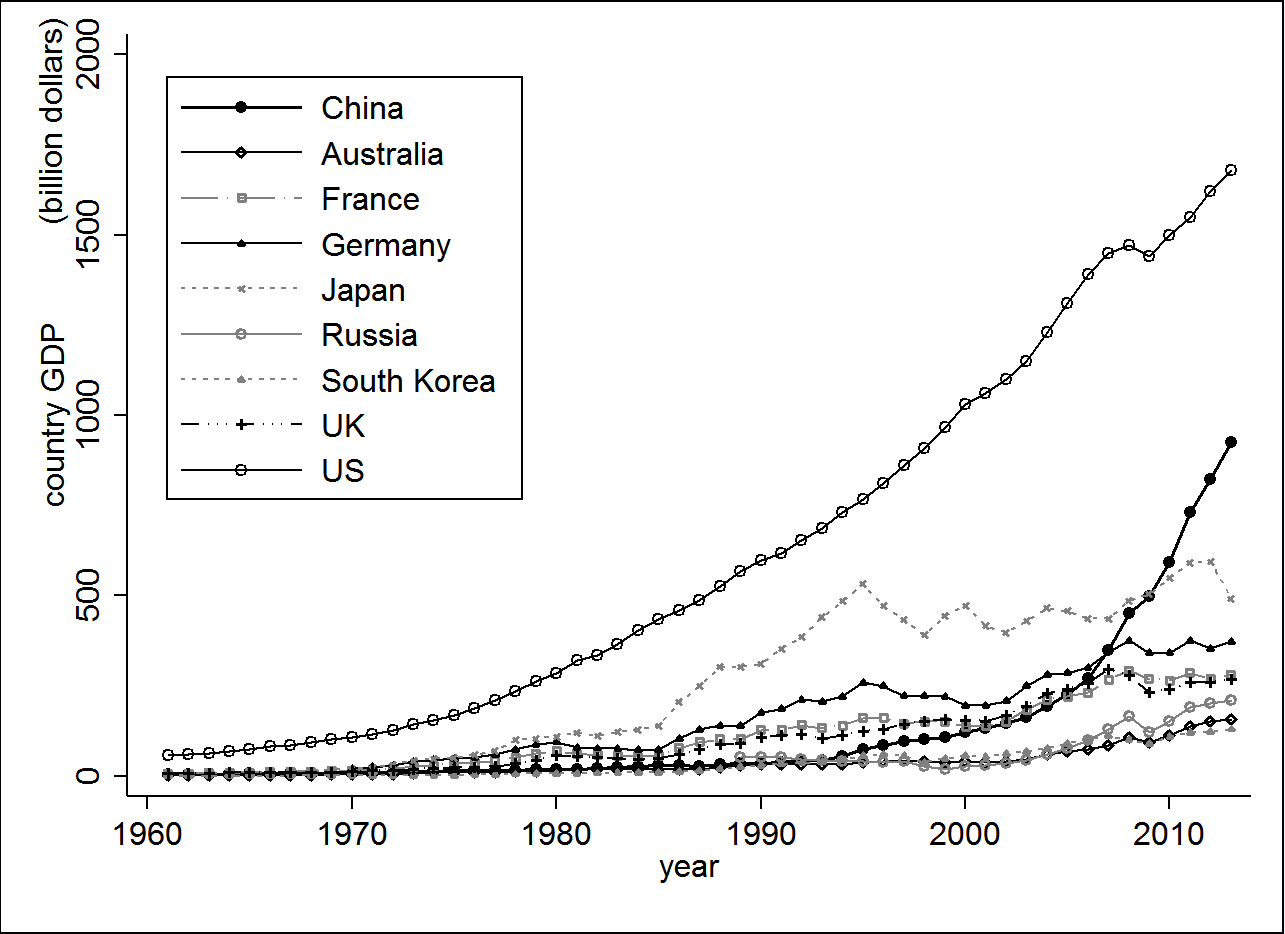}
\caption{China's spectacular economic growth in perspective (1961-2013).}
\end{figure}
The rest of the paper is organized as follows. Section 2 discusses related work in international relations, text classification and sentiment analysis. Section 3 describes the details of our dataset, the pre-processing procedures and the models. Section 4 presents the empirical findings. Section 5 discusses several interesting patterns in the context of international relations. Section 6 concludes this study.

\section{Related Work}
Our work is motivated by the observation that China's rise is rapidly reshaping the international structure. We follow Kenneth Waltz's definition that the international structure is defined by the arrangement of the states and that the structure changes with changes in the distribution of capabilities across the national states \cite{Waltz}. We concur with David Lake's argument that the international system is hierarchical and we contend that because of China's rapid rise, the relationship between China and many other states is now being redefined \cite{Lake}. Lake suggests that as China continues to grow, it is likely to form its own coalition of allies to counter America's current hierarchies.\footnote{An example will illustrate our point well. In March 2015, the UK became the first major Western country to seek to become a founding member of the Asian Infrastructure Investment Bank (AIIB), which is widely viewed as a rival to the US-dominated World Bank and the IMF. \textit{The Wall Street Journal} wrote an article describing UK's move as breaking ranks with Washington \cite{AIIB}.} Our work attempts to detect signs of realignment.
 
Our focus is on foreign perceptions of China's rise: whether the international community view it as a security threat or an economic opportunity \cite{Smith}. We believe foreign perceptions are important \cite{Jervis,Kydd}. In particular we are concerned that China's rise might be upsetting the sense of security of other nations. In this sense, our work seeks to empirically answer whether other nations believe China's rise can be peaceful or not, a question that has been theoretically analyzed by John Mearsheimer \cite{Mearsheimer}.

In order to understand and measure foreign perceptions, we examine foreign news media, which have long drawn the attention of IR scholars. For example, Ramos, Ron and Thoms study the reports of leading Western newspapers to answer the question of what influences the Northern media's coverage of events and abuses in explicit human rights terms \cite{Thoms}. Emilie Hafner-Burton constructs an autoregressive model to analyze the effects naming and shaming on political terror and political rights abuses \cite{Emilie}. More recently, Alastair Johnston analyzes Chinese publications and argues that China has not become more assertive despite all the Western suspicion and accusations \cite{Assertive}.

Methodologically, our work uses text classification and sentiment analysis. First we build a Naive Bayes classifier to investigate the evolving nature of foreign reports on China. This is related to many studies in text classification \cite{textasdata,Han}. In a recent study, a Naive Bayes classifier is used to classify SMS messages received by UNICEF Uganda into eleven classes \cite{voice}. The fact that their messages often contain various abbreviations and spelling errors seriously affects their initial results. The same data noise problem is not present here as we are examining the news articles published by leading news groups in respective countries. We test our classifier with 800 manually labeled articles, randomly chosen from the target newspapers. The test results, reported in Section 3, show that the Naive Bayes classifier is adequate for our task.

In addition to classifying news articles, our work also evaluates the sentiments of these articles. In related work, Pang \emph{et al.} evaluate the sentiments in movie reviews \cite{Cornell}. Agarwal \emph{et al.} analyze sentiments in Twitter messages \cite{agarwal:twitter}. Joo \emph{et al.} study the communicative intents of images \cite{ucla:image}. Compared with movie reviews and tweets, the news articles in this study are much longer.\footnote{As will be detailed later, for each newspaper we will concatenate all the articles published within a specific year into one super article. This makes our text long and rich.} The extraordinary lengths of these texts make them an ideal candidate for sentiment evaluation based on discriminant words. As we only have access to an English dictionary of negative words, when processing news articles in French and German, we first use Google Translate to translate that dictionary into French and German.

\section{Data and Methodology}
\subsection{Data Collection}
We have collected 132,834
 articles from 11 leading newspapers, all of which now have a web presence.\footnote{Our articles of \textit{The Wall Street Journal} come from ProQuest (http://search.proquest.com/advanced). All other articles come from LexisNexis (http://www.lexisnexis.com/hottopics/lnacademic).} Their summary statistics are provided in Table 1. First, for each country we select one representative newspaper and collect its news reports on China.\footnote{We do not claim that the selected newspapers are fully representative of national attitudes and we are aware that newspapers within a country may have conflicting views. But in terms of foreign policy, we believe views expressed by an influential newspaper offer a first-order approximation of the national attitudes.} The selected newspapers are \textit{The New York Times} (US), \textit{The Guardian} (UK), \textit{The Australian} (Australia), \textit{Le Figaro} (France), \textit{Die Welt} (Germany), \textit{The Daily Yomiuri} (Japan), \textit{RIA Novosti} (Russia) and \textit{Korea Times} (South Korea). \textit{The Daily Yomiuri}, \textit{RIA Novosti} and \textit{Korea Times} are published in English. \textit{Le Figaro} is in French; \textit{Die Welt} is in German.
\begin{table}[h]
\centering
\caption{Summary statistics of the news articles}
\begin{tabular}{llrl}\hline
Country     & Newspaper      & \# Articles & \multicolumn{1}{r}{Period}    \\
US          & New York Times & 21,411      & 1981-2014 \\
US          & Wall Street Journal & 32,111  & 1984-2014 \\
UK          & Guardian   & 11,130       & 1985-2014 \\
UK          & Telegraph  & 6,813       & 2001-2014 \\
UK          & Financial Times& 21,029      & 1982-2014 \\
AU   & The Australian & 11,071       & 1995-2014 \\
FR      & Le Figaro      & 7,990        & 1997-2014 \\
DE     & Die Welt       & 7,467        & 2001-2014 \\
JP       & Yomiuri        & 3,431        & 1990-2014 \\
RU      & RIA Novosti    & 5,853        & 2000-2014 \\
KR & Korea Times   & 4,528        & 1999-2014 \\
\hline
\end{tabular}
\end{table}

In collecting news articles, we use the keywords \textit{China} and \textit{Chinese} in the title search. We confirm that articles with keyword \textit{China's} as in \textit{China's cyber game} will also be collected. For the French-language newspaper, \textit{Le Figaro}, we use in our search the key word \textit{Chine} for \textit{China}, \textit{Chinois} for \textit{Chinese}. We confirm that articles with keyword \textit{Chinoise} as in \textit{la croissance chinoise} will also be collected. For the German-language newspaper, \textit{Die Welt}, we use the keywords \textit{China} and \textit{Chinesisch}. We confirm that articles with \textit{Chinesische}, \textit{Chinesischen}, and \textit{Chinesischer} in the headline are collected. 


Second, in order to examine the differences between a conservative newspaper and a liberal newspaper in the US and the UK, we collect articles from \textit{The Wall Street Journal} to compare with \textit{The New York Times}, and from \textit{The Daily Telegraph} (UK) to compare with \textit{The Guardian}.

Third, to build the Naive Bayes classifier, we obtain 972 labeled articles on China's economy and 995 labeled articles on Chinese politics from \textit{The Financial Times} (UK) through queries based on the title and subjects. From the labeled articles we extract 300 most common features for each class and create list$_{econ}$ and list$_{pols}$. The large corpus of the training data enables us to keep all the extracted features. For example, we are able to differentiate between \textit{Hong Kong} and \textit{Hong Kong's}. 

Lastly, we obtain 1,000 labeled articles on China's economy and 1,000 labeled articles on Chinese politics from \textit{Le Figaro} (France), and 600 labeled articles on China's economy and 600 labeled articles on Chinese politics from \textit{Die Welt} (Germany). In this way, we are able to train the classifier with three datasets in three  different languages. In the discussion that follows, we will focus primarily on the English-language classifier.
\subsection{Naive Bayes Classifier}

To save space, we do not report all the features in our calculation. They are available at the authors' website.\footnote{https://www.sites.google.com/site/wangyurochester.} Here we only give some examples. List$_{econ}$ contains such features as chines (2749), market (2679), bank (2428), govern (2188), growth (1886), compani (1798), econom (1791), shanghai (1761), invest (1761), develop (1160), trade (1008). List$_{pols}$ contains such features as parti (2713), chines (2582), hong kong (1845), polit (2151), govern (2588), beij (2198), protest (1699), japan (1189), democraci (1124), taiwan (735), militari (687). Notice that certain features may appear in both lists, such as \textit{econom} and \textit{govern}, but their frequencies differ.

To each feature $w_i$ in the news article we assign a weight in the following manner: 
\begin{equation}\phi(w_i)|list_j=\begin{cases}
    \frac{\text{count of w}_i\:\text{in list}_j\times 1000}{\text{total counts of list}_j},& \text{if}\:w_i\:\in\text{list}_j\\
    \underset{\text{w}_k\in \text{list}_j}{\min} \frac{\text{count of w}_k\:\text{in list}_j\times 1000}{\text{total counts of list}_j\times 10},& \text{otherwise;}
\end{cases}\end{equation}
where
\begin{equation}\underset{\text{w}_k\in \text{list}_{econ}}{\min}\frac{\text{count of w}_k\:\text{in list}_{econ}\times 1000}{\text{total counts of list}_{econ}\times 10}=0.0977,\end{equation}
\begin{equation}\underset{\text{w}_k\in \text{list}_{pols}}{\min}\frac{\text{count of w}_k\:\text{in list}_{pols}\times 1000}{\text{total counts of list}_{pols}\times 10}=0.1001.\end{equation}
If the feature is in the list, we assign to it a weight equal to the relative size of the feature in the list times 1000. If otherwise, we assign to it a uniform weight equal to the smallest relative size of any feature in the list times 100. Notice that here we have added a penalizing factor of 10 in the design. An article $l$ is then classified as economic if and only if\newline
\begin{equation}\prod_{\text{article}\:\textit{l}}{(\phi(w_i)|list_{econ})}>\prod_{\text{article}\:\textit{l}}{(\phi(w_i)|list_{pols})}.\end{equation}

That is, a score is calculated for each class as the product of the features' weights under the assumption of independence. The classifier then assigns the article to the class that gives it the higher score. In the (rare) case where the article contains none of the features in either list, it is classified as political by default. This is a result from (1)(2)(3)(4). Notice that here we have made the simplifying assumption that foreign reports on China are either economic or political: there is no third category such as culture or sports, and there is no overlap. 

We use 800 manually labeled articles to test the accuracy of the classifier. The test results are detailed in Table 2.
\begin{table}[h]
\centering
\caption{Test results for the news classifier}
\begin{tabular}{lll}\hline
Newspaper      & \# Articles & Accuracy \\
New York Times & 100         & 74\%    \\
The Guardian   & 100         & 83\%    \\
The Australian & 100         & 87\%    \\
Le Figaro      & 100         & 85\%    \\
Die Welt       & 100         & 78\%    \\
Yomiuri        & 100         & 84\%    \\
RIA Novosti    & 100         & 72\%    \\
Korea Times    & 100         & 73\%    \\
Average        & 100         & 80\%  \\
\hline
\end{tabular}
\end{table}
\subsection{Sentiment Analysis}
When evaluating the sentiments of the news articles, we use a collection of 4,790 negative words, 4,783 of which come from an established dictionary \cite{Illonois}.To this collection we have added 7 words that have strong negative connotations in China-related articles: military, dominate, authoritarian, communist, dictatorship, dalai (lama) and pollution.\footnote{We agree with Stephen Krasner that language usage reflects power distribution \cite{Krasner}.}

Since our focus is not on individual articles but on newspapers as a whole, sentiment analysis based on single articles would be less than perfect in this study. One strongly negative report could be more influential than two mildly positive reports, a phenomenon described in psychological studies as ``the positive-negative asymmetry effect'' \cite{Peeters:Negativity,roy:bad}. The example below helps illustrate our point.\footnote{The original texts come from \textit{The Financial Times}, January 14 and October 27, 2004.}

Excerpt 1. \textit{Specifically, the provinces will accelerate construction of roads and railways.
Guangdong intends to build a road link with the four neighbouring provinces by
next year, and a motorway to each area by 2007.} (Mildly Positive)

Excerpt 2. \textit{The region will also promote co-operation on energy, bringing power from the
less-developed west to the more industrialised east.} (Mildly Positive)

Excerpt 3. \textit{Meanwhile, widening income gaps between China's urban and rural areas and
coastal and interior provinces are likely to fuel social discontent while
endemic corruption undermines respect for a Communist regime that has very
little ideological legitimacy left.} (Strongly Negative)

We believe the overall evaluation of the three texts should be negative, but evaluating texts individually would lead us to conclude that the overall sentiment is positive. In order to overcome this problem, we define a holistic metric, the China Friendliness Index, to evaluate all the texts published by a newspaper within a year. The China Friendliness Index is constructed as follows:
\begin{equation}{\textit{I}}=25-\frac{300\times\text{(frequency of negative words)}}{\text{total length of news articles}}.\end{equation}

The number 300, which we arrived at through trial and error, is a scaling factor that makes sure the index stays between 0 and 25. Subtracting from 25 then transforms what is essentially a hostility measure into a measure of friendliness. We design the Index based on negative words because psychological studies have consistently shown that the psychological effects of bad ones outweigh those of the good ones \cite{Peeters:Negativity,roy:bad}.

\section{Empirical Results}
This section answers the following three questions one by one: Is the world talking more about China? What are other countries talking about regarding China: politics or economics? Are other countries feeling more or less positive towards China?
\subsection{Volume of the Coverage}
As China's economy continues to grow, so does its influence on the world stage. It is therefore natural to expect that China is now receiving more attention from foreign media than before. This hypothesis can be tested by calculating the absolute number of times that China appears in their news headlines and the percentage of China-focused articles for each newspaper each year. The result is shown in Figure 2.\footnote{Data on \textit{The Australian} for the year 2005 are not complete.} To test for significance, we regress the percentage of China-focused articles against year. The regression coefficients (coef.) and t statistics (t) are reported following Figure 2.
\begin{figure}[h!]
\centering
\includegraphics[width=8.9cm, height=6.5cm]{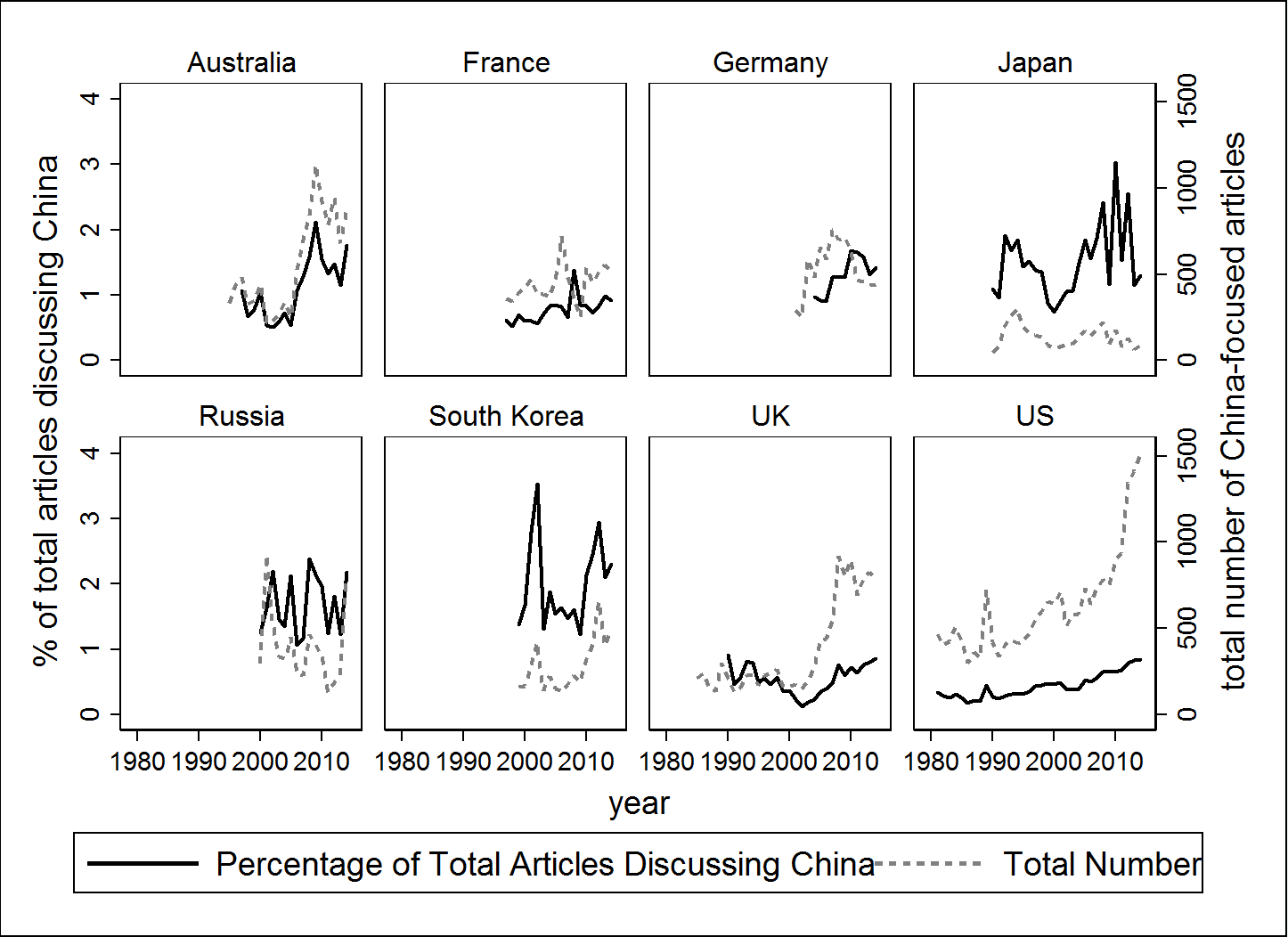}

\vspace{0.8cm}
\begin{tabular}{rlll}
\hline
\textbf{Australia} & coef.=0.060  & t=3.72  & \textbf{p=0.002} \\
\textbf{France}    & coef.=0.023  & t=3.11  & \textbf{p=0.007} \\
\textbf{Germany}   & coef.=0.065  & t=3.67  & \textbf{p=0.005} \\
Japan              & coef.=0.023 & t=1.57  & p=0.130          \\
Russia             & coef.=0.016  & t=0.57  & p=0.579          \\
S. Korea           & coef.=0.020  & t=0.55  & p=0.588          \\
UK                 & coef.=0.003  & t=0.46  & p=0.650          \\
\textbf{US}        & coef.=0.016  & t=10.52 & \textbf{p=0.000} \\
\hline
\end{tabular}
\vspace{.5cm}
\captionsetup{justification=justified,
singlelinecheck=false
}
\caption{Is the world talking more about China?}
\end{figure}

The result shows that all the countries under study have been reporting more on China as measured by the percentage of China-focused articles. The increase is statistically significant for Australia, France, Germany and the US. We also report that as measured by the time series average, South Korea (2.00\%), Russia (1.80\%) and Japan (1.51\%) report on China most frequently.
\subsection{Content of the Coverage}
China's rise is first and foremost an economic rise. Chinese politics has also changed in the same period, but not nearly by the same magnitude. Therefore, one would expect that the world's interest in China has been shifting towards China's economy and away from Chinese politics. This is where the Naive Bayes classifier can help. For each year and each newspaper, we calculate the percentage of China-centered articles that are economic. To test for significance, we regress the percentage variable against time. The regression coefficients and t-statistics are reported following Figure 3.

\begin{figure}[h!]
\centering
\includegraphics[width=8.9cm, height=6.5cm]{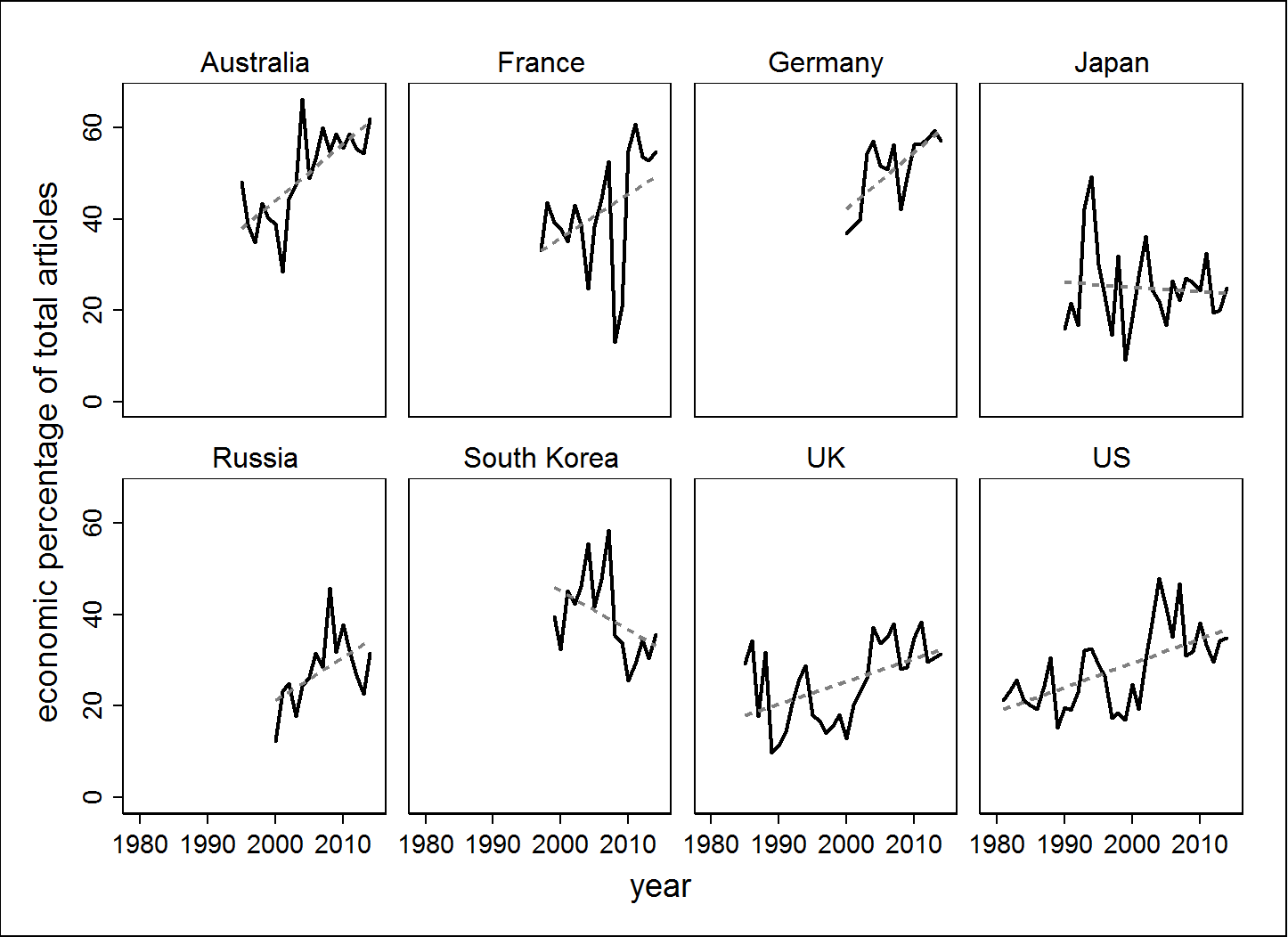}
\centering

\vspace{.8cm}
\begin{tabular}{rlll}
\hline
\textbf{Australia} & coef.=1.232  & t=4.57  & \textbf{p=0.000} \\
France    & coef.=0.956  & t=1.74  & p=0.102 \\
 \textbf{Germany}   & coef.=0.064  & t=3.49  & \textbf{p=0.007} \\
Japan              & coef.=-0.099 & t=-0.40 & p=0.694         \\
\textbf{Russia}    & coef.=0.939  & t=2.2   & \textbf{p=0.045} \\
\textbf{S. Korea}  & coef.=-0.856 & t=-1.83 & \textbf{p=0.089} \\
\textbf{UK}        & coef.=0.496  & t=3.06  & \textbf{p=0.005} \\
\textbf{US}        & coef.=0.530  & t=4.45  & \textbf{p=0.000}\\
\hline
\end{tabular}
\vspace{.5cm}
\captionsetup{justification=justified,
singlelinecheck=false
}
\caption{What is the world talking about regarding China?}
\end{figure}

We find that the majority of the countries are focusing more on China's economy and less on Chinese politics and that in 2008 there is a dip in economic reporting in all the five Western countries. The dip is particularly sharp in France. Protests in Paris against the Beijing Olympic Games and the then French president Sarkozy's decision to meet with Dalai Lama severely strained the two countries' relations and led to the so-called ``Chinese-French rift'' \cite{eubusiness}. We also find that reporting in Japan and South Korea is shifting towards Chinese politics. The shift is statistically significant for South Korea. China is the largest trading partner for both countries and yet both seem to be talking more about Chinese politics. One plausible explanation for their behavior is geopolitics. Both are China's close neighbors and both are allies of the United States. They might perceive China's rise as a security threat. This can be better analyzed in terms of sentiment of the coverage, which is the topic of the next subsection.
\subsection{Sentiment of the Coverage}
For each newspaper and each year, we collect all its articles on China, concatenate them, and calculate its overall China Friendliness Index. The result is shown in Figure 4. To test the significance of the trend, we regress the Index against time for each country. Regression results are also reported.

\begin{figure}[h!]
\centering
\includegraphics[width=8.9cm, height=6.5cm]{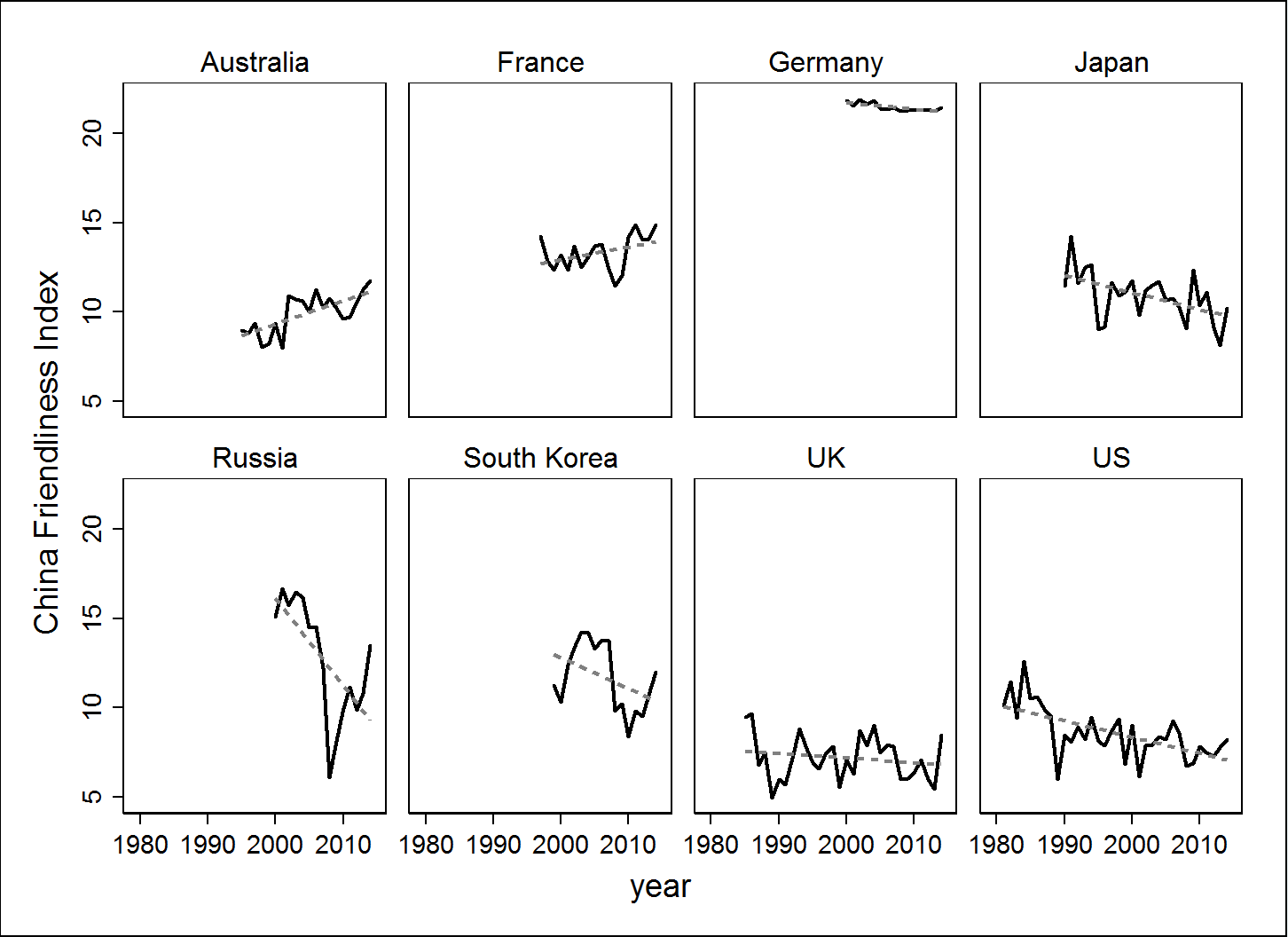}

\vspace{.8cm}

\begin{tabular}{rlll}
\hline
\textbf{Australia} & coef.=0.058  & t=3.03  & \textbf{p=0.007} \\
France             & coef.=0.013  & t=0.68  & p=0.507          \\\textbf{Germany}   & coef.=-0.034  & t=-5.21 & \textbf{p=0.000}
\\

\textbf{Japan}     & coef.=-0.044 & t=-2.61 & \textbf{p=0.016}  \\
\textbf{Russia}    & coef.=-0.259 & t=-4.02 & \textbf{p=0.001} \\
S. Korea           & coef.=-0.017 & t=-0.47 & p=0.648          \\
UK                 & coef.=-0.014 & t=-1.22 & p=0.233         \\
\textbf{US}        & coef.=-0.037 & t=-3.89 & \textbf{p=0.000}\\
\hline
\end{tabular}

\vspace{.5cm}
\captionsetup{justification=justified,
singlelinecheck=false
}
\caption{How is the world feeling about China: positive or negative?}
\end{figure}

According to the data, Australia is the only country under examination that has become unequivocally more friendly towards China. Indeed, Australia is the only country that we identify as realigning with China: it is reporting more on China, the reports are increasingly focused on China's economy, and it is becoming more positive towards China.\footnote{Huntington has argued that Australia is a ``torn'' country where its leaders try to delink their country from the West and make it a part of Asia \cite{Huntington:Civilization}.} The China Friendliness Index in France has risen too, but the increase is not statistically significant. The Friendliness Index has dropped in all other countries. The drop is statistically significant in Germany, Japan, Russia and the US.

For external validation, we compare our Index with public opinion poll data. As noted, time series poll data on attitudes towards China is rare; only Gallup (US) has opinion data that goes back to 1979, and Genron-NPO (Japan) has opinion data that covers the period between 2005 and 2014. In order to use a consistent source, we will use survey data from Pew Research Center, a widely recognized data source and authority for international relations studies \cite{nye,eastasia}. Pew's surveys, from 2007 to 2013, cover seven of the eight countries under examination, excepting only Australia, which was surveyed only twice in the period.\footnote{http://www.pewresearch.org.} For the case of Australia, we decide to use data from Lowy Institute for International Policy, which covers all the years between 2006 and 2014.\footnote{http://www.lowyinstitute.org.} The result is shown in Figure 5.\footnote{In completing Figure 5, we first apply an affine transformation to the survey data. The transformation does not affect the correlation coefficient, as corr(x, ay+b) = corr(x,y).} The correlation coefficient (corr) for each country is also reported.
\begin{figure}[h]
\centering
\includegraphics[width=8.9cm, height=6.5cm]{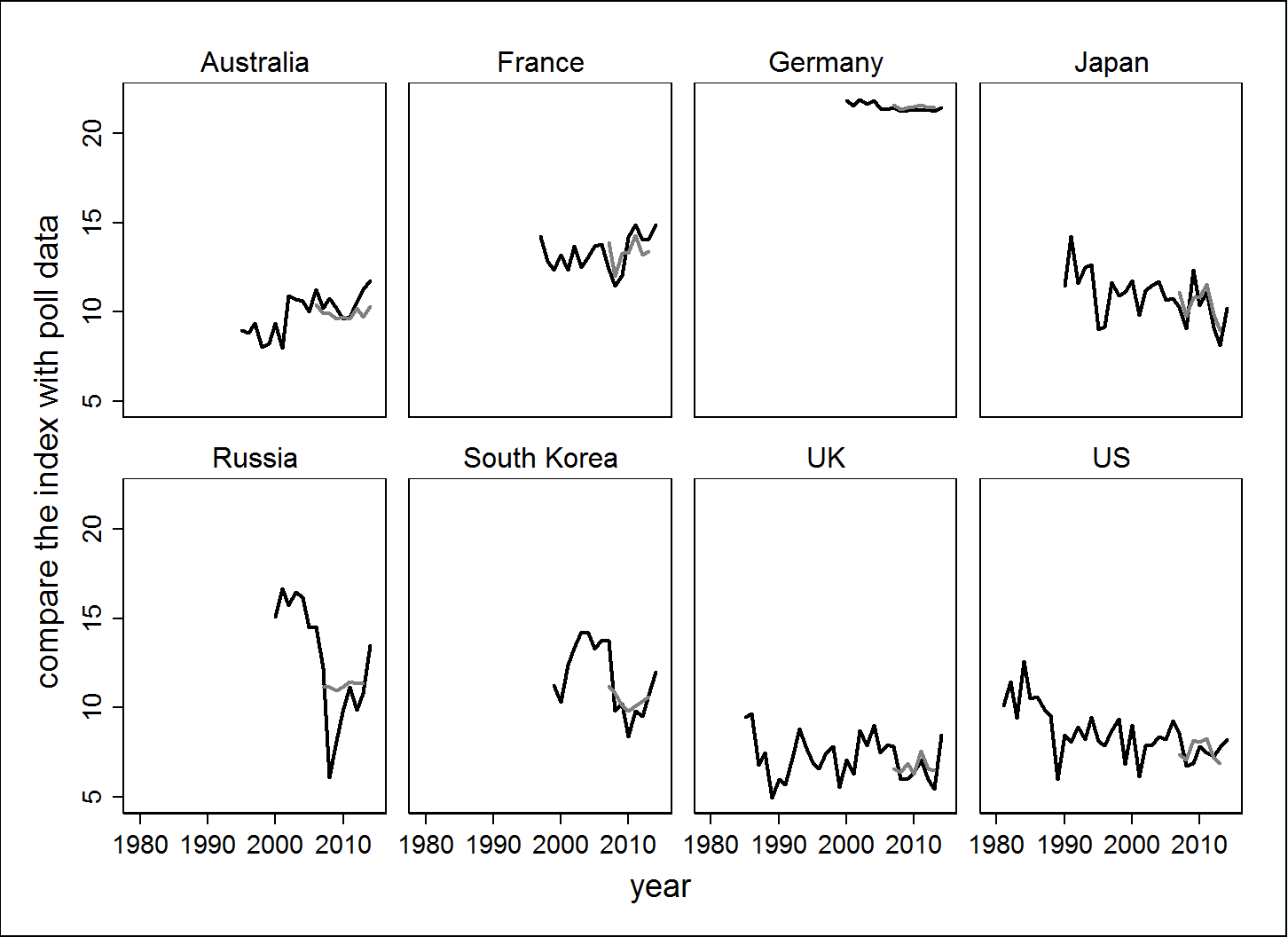}

\vspace{.8cm}
\begin{tabular}{rllrl}
\hline
\textbf{Australia} & \textbf{corr=0.69} &  & Russia            & corr=0.48          \\
\textbf{France}    & \textbf{corr=0.62} &  & \textbf{S. Korea} & \textbf{corr=0.84} \\
\textbf{Japan}     & \textbf{corr=0.83}  &  & UK                & corr=0.37          \\
\textbf{Germany}   & \textbf{corr=0.69}  &  & US                & corr=-0.06        \\
\hline
\end{tabular}

\vspace{.5cm}
\captionsetup{justification=justified,
singlelinecheck=false
}
\caption{Compare the China Friendliness Index with opinion poll data.}
\end{figure}
 

Comparing the Index with survey data from Pew Research Center, the highest correlations are achieved in South Korea (0.84) and Japan (0.83) and the lowest two are in the UK (0.37) and the US (-0.06). Given that opinion polls are carried out only in certain periods of a year, while the Index is constructed on an annual basis, the correlation analysis is satisfactory. It should be noticed, however, that using a different set of survey data may produce different results. For example, the correlation coefficient between the China Friendliness Index and Gallup opinion poll data on the US is 0.6834.\footnote{http://www.gallup.com/poll/1627/China.aspx.}

While we do expect our Index to be positively correlated with opinion poll data, we suggest that lower correlation coefficients as in the case of Russia and the US not be interpreted as inferior results. A lower correlation could reflect a stronger role of the news media in agenda-setting \cite{ucla:image}. Ideally we should compare our results with the aggregated poll data from different pollsters to eliminate potential bias, but data on attitudes towards China is generally lacking. 

\section{China's Bilateral Relations: Detailed Analysis}
In this section, we present findings on four topics that are of particular importance to China. 1). Conservative newspapers are more China-friendly than the liberal ones. 2). The special relationship between the US and the UK extends to their foreign relations with China. 3). ``Hot economics and cold politics'' is a general feature of China's bilateral relations. 4). Hong Kong protests do not have a significant impact on China's foreign relations. 

\subsection{Conservative vs Liberal}
Like political parties, newspapers can be more conservative (right) or liberal (left). Is there any difference in their reports on China? To answer this question, we contrast the sentiments towards China in left and right newspapers in the US and the UK. For the US, we compare \textit{The Wall Street Journal} (right) and \textit{The New York Times} (left); for the UK, we study \textit{The Telegraph} (right) and \textit{The Guardian} (left). The result is shown in Figure 6. 

\begin{figure}[h!]
\centering
\includegraphics[width=8.9cm, height=6.5cm]{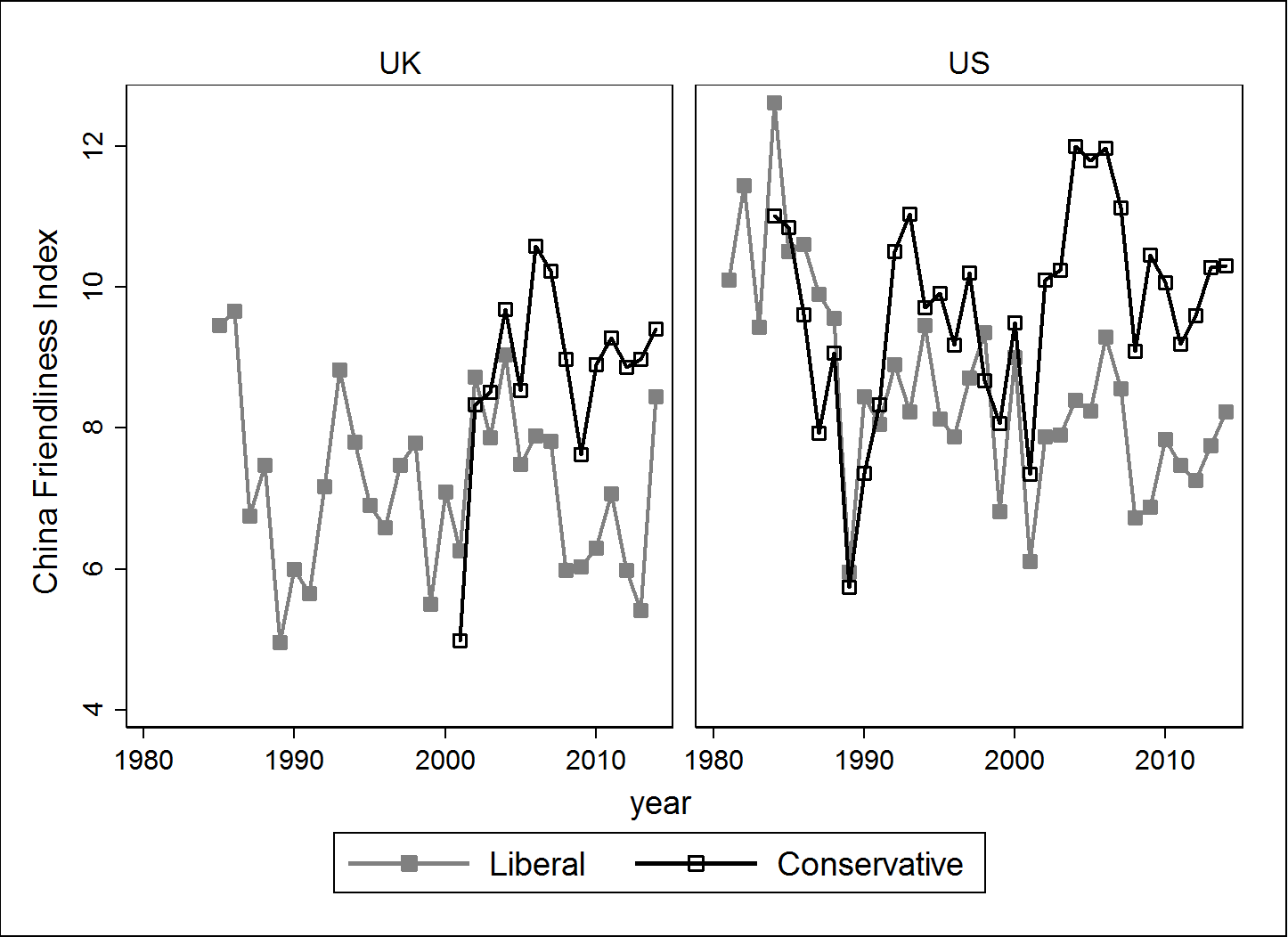}
\captionsetup{justification=justified,
singlelinecheck=false
}
\caption{Conservative newspapers are more China-friendly than liberal newspapers.}

\end{figure}

The results show that the China Friendliness Index is higher for the conservative newspaper in both countries. In the US, sentiments reported by the conservative newspaper and the liberal newspaper have diverged substantially over the past two decades. This is exactly the period when China's economy took off. Assuming that the difference between the two newspapers' index values is normally distributed between 1991 and 2014, we perform a paired t test for statistical significance. We find that the difference between the two newspapers is statistically significant.  

\begin{table}[h!]
\centering
\caption{Paired t test (Wall Street Journal and New York Times)}
\begin{tabular}{lllllll}
\multicolumn{1}{l|}{Variable} & Obs & Mean     & Std. Err.                & Std. Dev.                  & {[}95\% Conf.            & Interval{]}            \\ \hline
\multicolumn{1}{l|}{right}    & 24  & 9.945 & 0.241 & 1.180                   & 9.447                & 10.444               \\
\multicolumn{1}{l|}{left}     & 24  & 8.047 & 0.177                 & 0.868         & 7.681                  & 8.414               \\ \hline
\multicolumn{1}{l|}{diff}     & 24  & 1.898 & 0.221                 & 1.084                   & 1.440                 & 2.356               \\ \hline
\multicolumn{4}{l}{mean(diff) = mean(right - left)}                       & \multicolumn{3}{r}{t = 8.799}                                                \\
\multicolumn{4}{l}{Ho: mean(diff) = 0}                                    & \multicolumn{3}{r}{degrees of freedom = 23}                                    \\
\multicolumn{3}{l}{Ha: mean(diff) \textless 0} & \multicolumn{2}{c}{Ha: mean(diff) != 0}               & \multicolumn{2}{r}{Ha: mean(diff) \textgreater 0} \\
\multicolumn{3}{l}{Pr(T \textless t) = 1.000} & \multicolumn{2}{c}{Pr( $\vert T\vert$ \textgreater t) = 0.000} & \multicolumn{2}{r}{Pr(T \textgreater t) = 0.000}
\end{tabular}
\end{table}
Comparison in the UK is limited by data availability but we can see that after 2002, the conservative \textit{Telegraph} has invariably been more friendly towards China than the more liberal \textit{Guardian}. We perform the same paired t-test on the Index scores for \textit{The Telegraph} and \textit{The Guardian} from 2003 to 2014. We find that the difference between the two newspapers is significant. One possible explanation for this disparity is that conservative newspapers stand for big businesses and free trade, and thus tend to view China more in terms of economic opportunities.
\begin{table}[h]
\centering
\caption{Paired t test (The Telegraph and The Guardian)}
\begin{tabular}{lllllll}
\multicolumn{1}{l|}{Variable} & Obs & Mean     & Std. Err.                & Std. Dev.                  & {[}95\% Conf.            & Interval{]}            \\ \hline
\multicolumn{1}{l|}{right}    & 12  & 9.132 & 0.228                & 0.789                  & 8.632                  & 9.633               \\
\multicolumn{1}{l|}{left}     & 12  &  7.112  & 0.332 &  1.150  & 6.381                  &7.843              \\ \hline
\multicolumn{1}{l|}{diff}     & 12  & 2.020 & 0.289                  & 1.001                    & 1.384                & 2.657 \\ \hline
\multicolumn{4}{l}{mean(diff) = mean(right - left)}                       & \multicolumn{3}{r}{t =  6.988}                                                \\
\multicolumn{4}{l}{Ho: mean(diff) = 0}                                    & \multicolumn{3}{r}{degrees of freedom = 11}                                    \\
\multicolumn{3}{l}{Ha: mean(diff) \textless 0} & \multicolumn{2}{c}{Ha: mean(diff) != 0}               & \multicolumn{2}{r}{Ha: mean(diff) \textgreater 0} \\
\multicolumn{3}{l}{Pr(T \textless t) = 1.000} & \multicolumn{2}{c}{Pr( $\vert T\vert$ \textgreater t) = 0.000} & \multicolumn{2}{r}{Pr(T \textgreater t) = 0.000}
\end{tabular}
\end{table}  

\subsection{Correlation between the UK and US Reporting}
``Special relationship'' is a term used to characterize the close economic, cultural and political ties between the United Kingdom and the United States \cite{UK-US}. A closer analysis reveals that this special relationship also extends to their attitudes towards China. Trends in the two countries' China Friendliness Index closely match each other, as evidenced in Figure 7.

\begin{figure}[h]
\centering
\includegraphics[height=6.5cm, width=8.9cm]{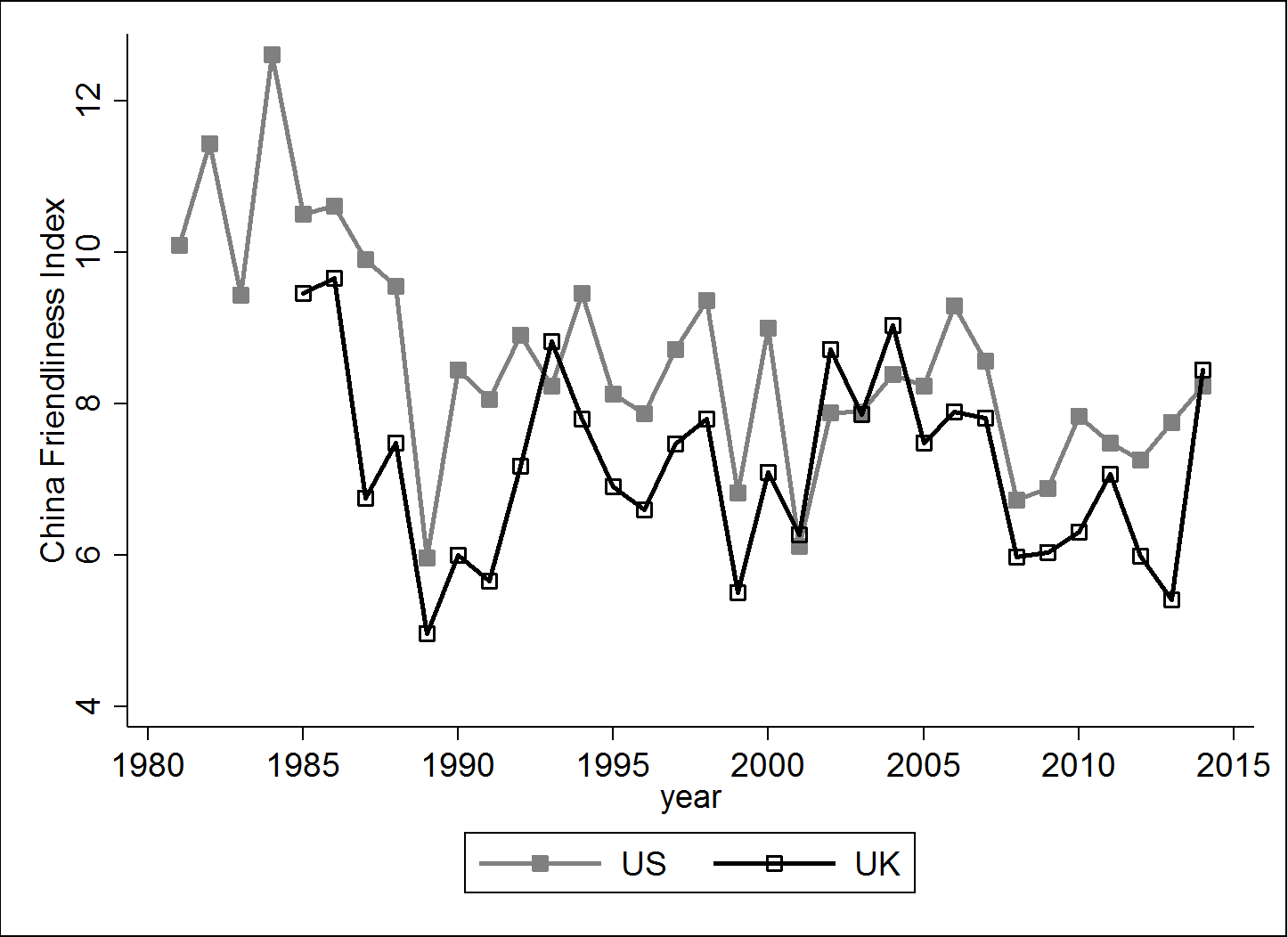}
\captionsetup{justification=justified,
singlelinecheck=false
}
\caption{UK and US, united in their attitudes towards China.}
\end{figure}

Moreover, by calculating correlation coefficients of the indexes, we are able to show that the correlation between the UK and the US is among the highest of all the 28 country pairs. We notice that the strongest correlation occurs between Germany and Russia, which we did not anticipate. But this very discovery draws our attention to the existence of a rich literature on the ``special relationship" between Germany and Russia \cite{Alexander:Germany}. We also point out that the observation period for the UK-US relationship (30 years) is substantially longer than that for the Germany-Russia relationship (14 years).
\begin{table}[h]
\tabcolsep=0.16cm
\centering
\caption{Correlations in Friendly Reporting on China}
\begin{tabular}{lllllllll}\hline
   & US  & UK  & FR  & DE  & AU     & KR  & JP  & RU \\
US & 1    &      &      &      &         &      &      &    \\
UK & \textbf{0.68} & 1    &      &      &         &      &      &    \\
FR & 0.35 & 0.27 & 1    &      &         &      &      &    \\
DE & 0.33 & \textbf{0.67} & -0.06 & 1    &         &      &      &    \\
AU & 0.15 & 0.31 & 0.31 & -0.09 & 1       &      &      &    \\
KR & 0.41 & \textbf{0.72} & -0.16 & 0.52 & 0.22     & 1    &      &    \\
JP & 0.33 & 0.26 & -0.10 & 0.50 & -0.07    & 0.28 & 1    &    \\
RU & 0.40 & \textbf{0.67} & 0.13 & \textbf{0.77} & -0.15 & \textbf{0.75} & 0.38 & 1 \\
\hline 
\end{tabular}
\end{table}

\subsection{Cold Politics and Hot Economics}
``Cold politics and hot economics,'' often regarded as a defining feature of China-Japan relations, refers to the fact that China and Japan trade heavily with each other and yet both have a very unfavorable view of each other when it comes to foreign relations \cite{Koo:Japan}. Our study explores whether or not ``cold politics and hot economics'' applies, in different degrees, to China's other bilateral relations. As described in Section 3, we first classify the news stories as either economic or political, then concatenate the articles in each group, and lastly calculate the China Friendliness Index for each group. The result is shown in Figure 8. We also conduct paired t-tests, reported beneath Figure 8, to compare the index for economic articles and the index for political articles. We find that the difference is statistically significant for all the eight countries.\newline
\begin{figure}[h]
\centering
\includegraphics[height=6.5cm, width=8.9cm]{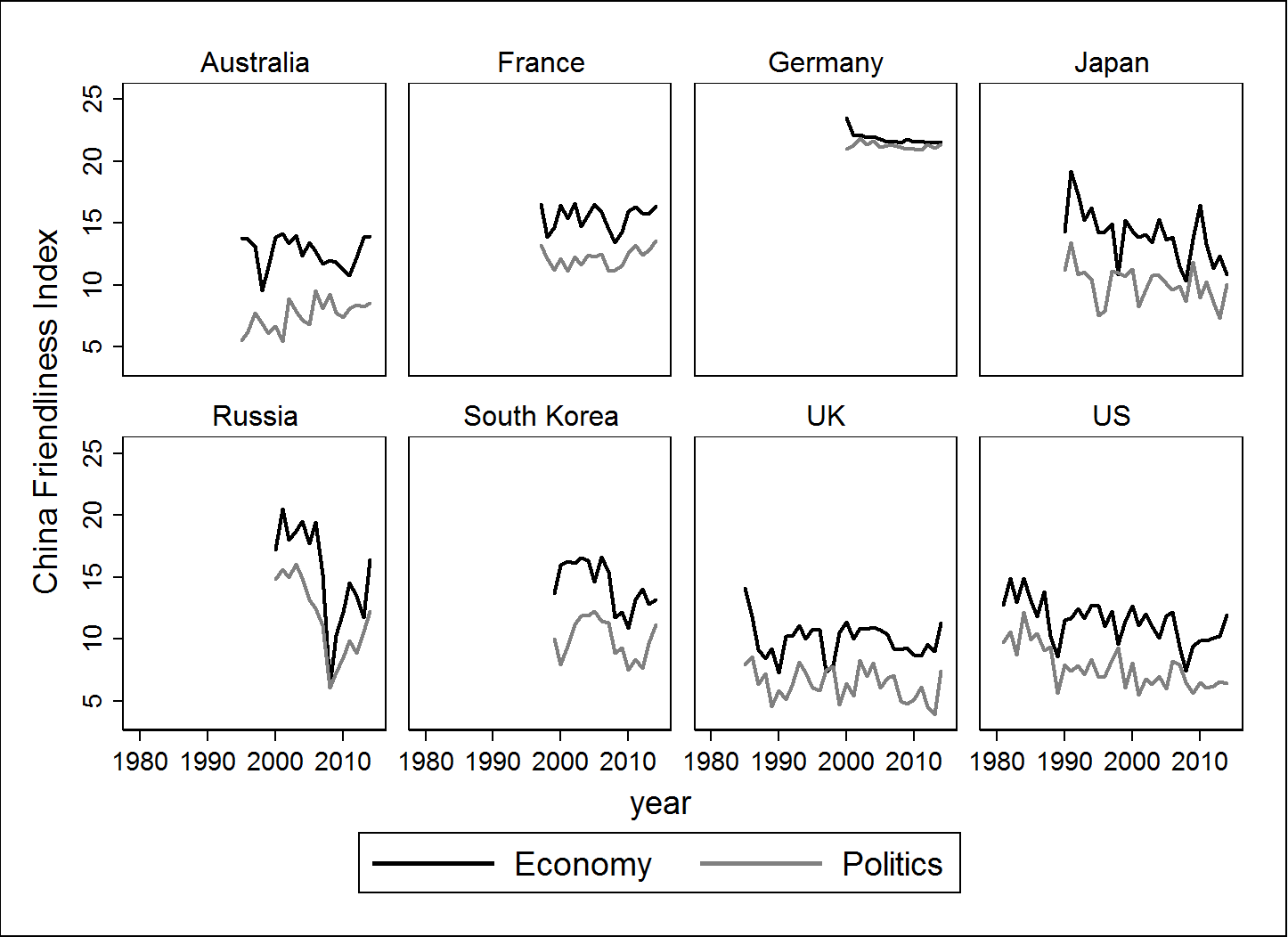}

\vspace{.5cm}

\begin{tabular}{rlll}
\hline
\textbf{Australia} & mean=5.11 & t=12.37 & \textbf{p=0.000} \\
\textbf{France}    & mean=3.28 & t=19.99 & \textbf{p=0.000} \\
\textbf{Germany}   & mean=0.61 & t=4.09  & \textbf{p=0.001} \\
\textbf{Japan}     & mean=3.94 & t=10.09 & \textbf{p=0.000} \\
\textbf{Russia}    & mean=3.63 & t=8.38  & \textbf{p=0.000} \\
\textbf{S. Korea}  & mean=4.37 & t=10.41 & \textbf{p=0.000} \\
\textbf{UK}        & mean=3.58 & t=12.53 & \textbf{p=0.000} \\
\textbf{US}        & mean=3.76 & t=15.29 & \textbf{p=0.000}\\
\hline
\end{tabular}

\vspace{.5cm}

\captionsetup{justification=raggedright,
singlelinecheck=false
}

\caption{Cold politics and hot economics.}
\end{figure}
\: The findings are quite surprising. Apparently ``cold politics and hot economics'' applies to all of China's bilateral relations under examination. Further research is needed to decide whether ``cold politics and hot economics'' is a general description of international relations, of which China-Japan relations are just an extreme case, or it applies only to China and China's bilateral relations. 
\subsection{Impact of 2014 Hong Kong protests}
Large-scale protests against Beijing's proposed electoral reforms erupted in Hong Kong in September 2014. The protests had received continuous and intensive coverage by worldwide media, and were described by \textit{The Wall Street Journal} as Hong Kong's ``most serious confrontation with Beijing in more than a decade'' \cite{wsj:hongkong}. It is therefore important to be able to measure their real impact. On the economic front, the World Bank has answered the question.  The event ``does not appear to have an impact on the overall business confidence'' \cite{world_bank}. This study examines the political front and shows that the event does not appear to have an impact on China's bilateral relations.



\begin{figure}[h!]
\centering
\includegraphics[height=6.5cm, width=8.9cm]{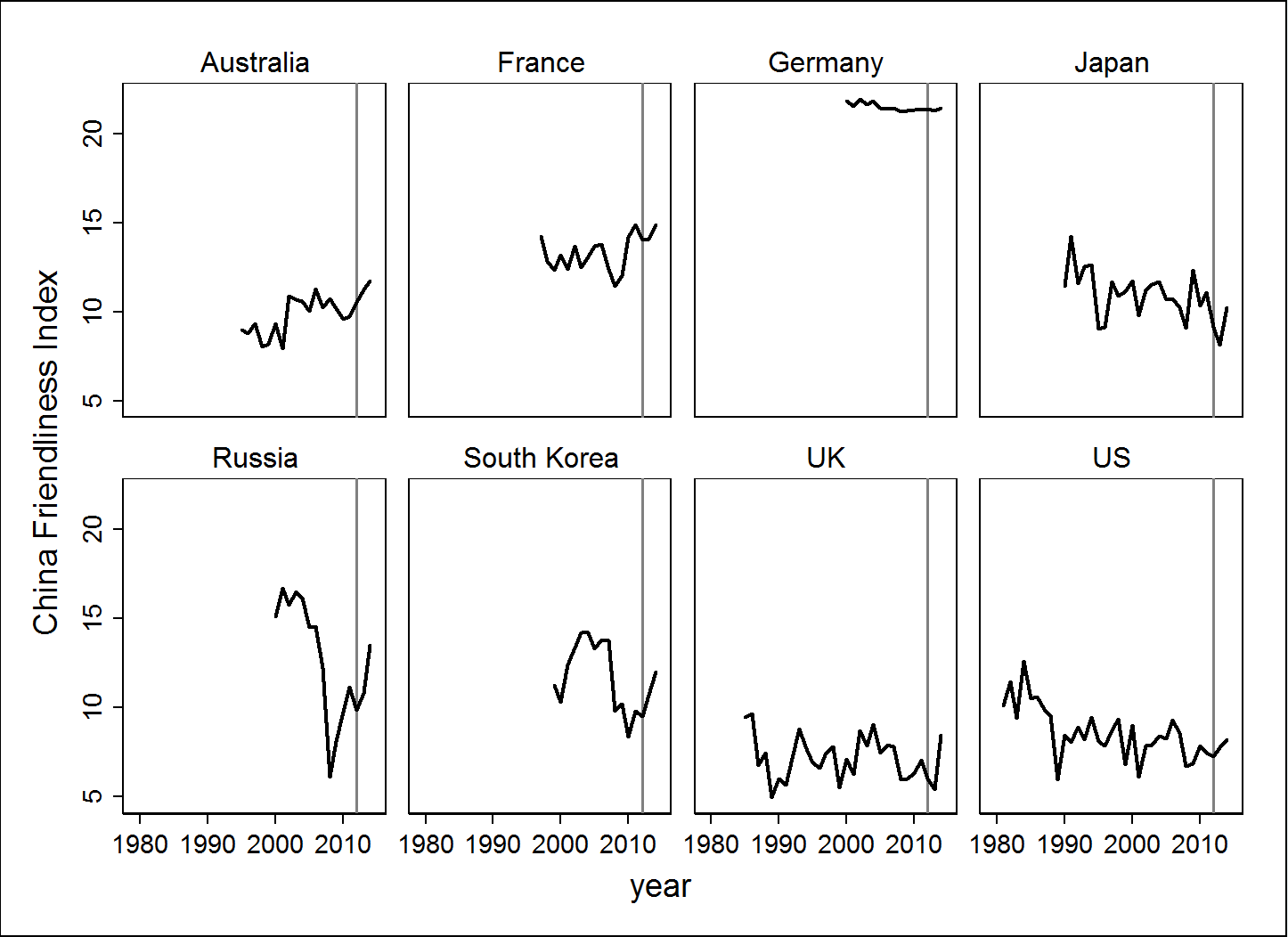}
\caption{Put the Hong Kong protests in perspective.}
\end{figure}

As Figure 9 shows, the Friendliness Index has actually improved for all the eight countries in 2014.\footnote{Simultaneous warming in the eight countries has historical precedents. For example, the eight countries became unanimously more friendly towards China in 2002.} To better evaluate the event's impact, we carry out a cross-sectional analysis on a monthly basis. Here we study the UK, which ruled Hong Kong until 1997, and the US. The findings are reported in Figure 10.

\begin{figure}[h!]
\centering
\includegraphics[width=8.9cm, height=5.5cm]{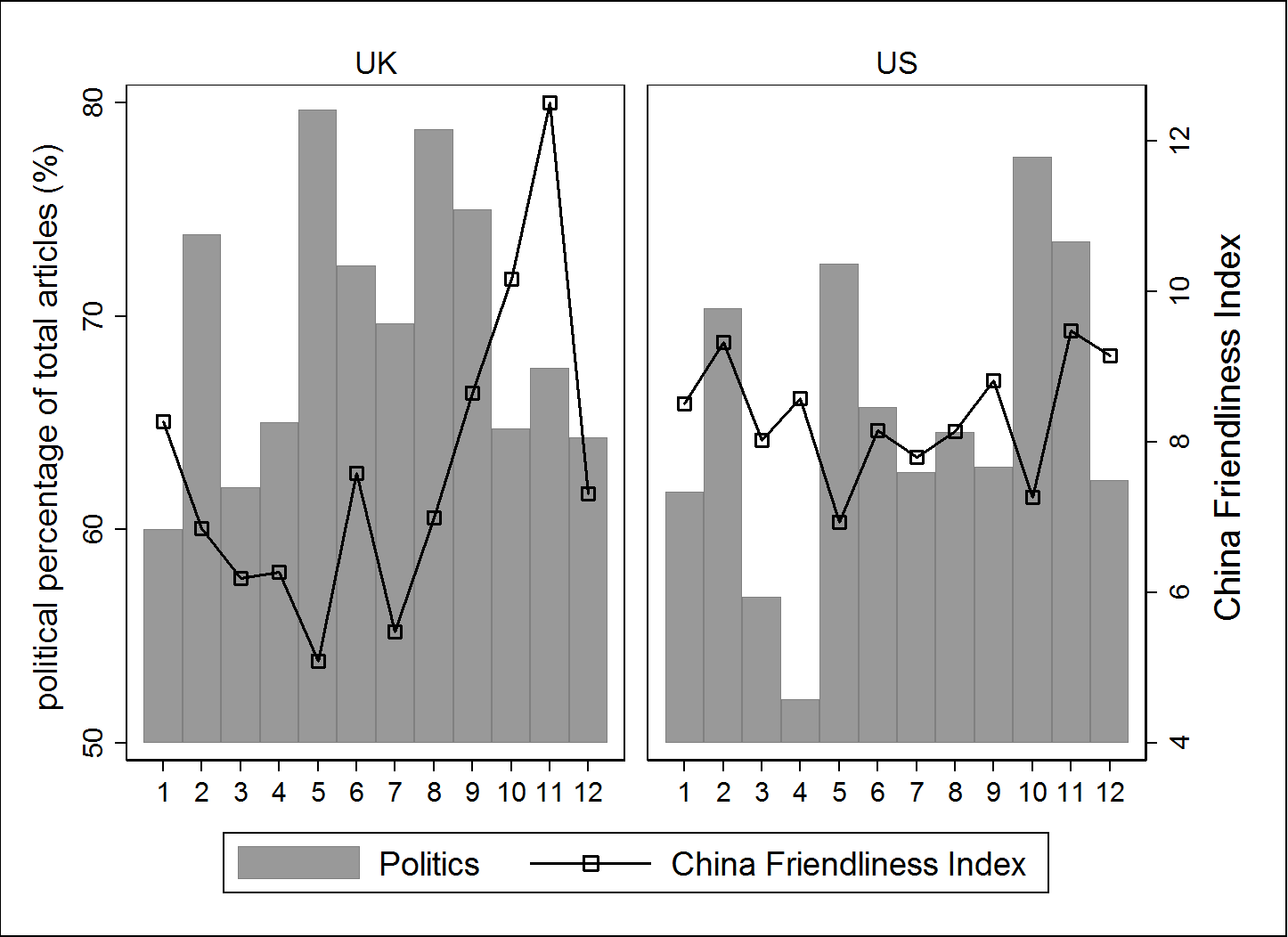}
\caption{Effects of Hong Kong protests on the Friendliness Index.}
\end{figure}

It can be seen that the protests, which started in late September, did not have a significant impact on the Index for that month in either country. In October, a dip in the Index was observed in the US and news reporting became more political, but the Index quickly recovered in November. No parallel dip was observed in the UK and there was no increase in political reporting. \textit{The Economist} later described Britain's response as ``limp'' \cite{abridgenottoofar}.


\section{Conclusions}
China's rise is rapidly changing the international structure, bringing both opportunities and challenges to other countries. International perceptions matter. In this article, we apply text classification and sentiment analysis techniques to capture these perceptions. 

We find that there has been increasing news coverage of China in all the countries under study. We also find that the emphasis of the foreign reports is generally shifting towards China's economy. Japan and South Korea are exceptions: they are reporting more on Chinese politics instead. In terms of global sentiment, the picture is quite gloomy. With the exception of Australia and, to some extent, France, all the other countries under examination are becoming less positive towards China. 

So far as we know, our paper is the first of its kind that applies data mining to the study of international relations based on large-scale online news media data. The results we achieve are very encouraging and they contribute to a better understanding of the evolving nature of China's bilateral relations.
\section*{Acknowledgment}
Yu Wang would like to thank the Department of Political Science at the University of Rochester for warm encouragement and generous funding. This work was also generously supported in part by Google, Yahoo,  Adobe, TCL, and New York State CoE CEIS and IDS.



\bibliographystyle{IEEEtran}
%

\bibliography{dream}

\begin{thebibliography}{10}
\providecommand{\url}[1]{#1}
\csname url@samestyle\endcsname
\providecommand{\newblock}{\relax}
\providecommand{\bibinfo}[2]{#2}
\providecommand{\BIBentrySTDinterwordspacing}{\spaceskip=0pt\relax}
\providecommand{\BIBentryALTinterwordstretchfactor}{4}
\providecommand{\BIBentryALTinterwordspacing}{\spaceskip=\fontdimen2\font plus
\BIBentryALTinterwordstretchfactor\fontdimen3\font minus
  \fontdimen4\font\relax}
\providecommand{\BIBforeignlanguage}[2]{{%
\expandafter\ifx\csname l@#1\endcsname\relax
\typeout{** WARNING: IEEEtran.bst: No hyphenation pattern has been}%
\typeout{** loaded for the language `#1'. Using the pattern for}%
\typeout{** the default language instead.}%
\else
\language=\csname l@#1\endcsname
\fi
#2}}
\providecommand{\BIBdecl}{\relax}
\BIBdecl

\bibitem{Waltz}
K.~N. Waltz, \emph{Theory of International Politics}.\hskip 1em plus 0.5em
  minus 0.4em\relax McGraw-Hill Publishing Company, 1979.

\bibitem{Lake}
D.~A. Lake, \emph{Hierarchy in International Relations}.\hskip 1em plus 0.5em
  minus 0.4em\relax Cornell University Press, 2009.

\bibitem{AIIB}
\BIBentryALTinterwordspacing
M.~Magnier. (2015, March) U.k. to join china-backed development bank. [Online].
  Available:
  \url{http://www.wsj.com/articles/u-k-to-join-china-backed-development-bank-1426211662}
\BIBentrySTDinterwordspacing

\bibitem{Smith}
S.~A. Smith, \emph{Intimate Rivals: Japanese Domestic Politics and a Rising
  China}.\hskip 1em plus 0.5em minus 0.4em\relax Columbia University Press,
  2015.

\bibitem{Jervis}
R.~Jervis, \emph{Perceptions and Misperception in International
  Politics}.\hskip 1em plus 0.5em minus 0.4em\relax Princeton University Press,
  1976.

\bibitem{Kydd}
A.~H. Kydd, \emph{Trust and Mistrust in International Relations}.\hskip 1em
  plus 0.5em minus 0.4em\relax Princeton University Press, 2005.

\bibitem{Mearsheimer}
J.~J. Mearsheimer, \emph{The Tragedy of Great Power Politics}.\hskip 1em plus
  0.5em minus 0.4em\relax W. W. Norton \& Compnay, 2014.

\bibitem{Thoms}
J.~R. Howards~Ramos and O.~N.~T. Thoms, ``Shaping the northern media's human
  rights coverage,'' \emph{Journal of Peace Research}, vol.~44, no.~4, pp.
  385--406, 2007.

\bibitem{Emilie}
E.~M. Hafner-Burton, ``Sticks and stones: Naming and shaming the human rights
  enforcement problem,'' \emph{International Organization}, vol.~62, no.~4, pp.
  689--716, 2008.

\bibitem{Assertive}
A.~I. Johnson, ``How new and assertive is china's new assertiveness?''
  \emph{International Security}, vol.~37, no.~4, pp. 7--48, 2013.

\bibitem{textasdata}
J.~Grimmer and B.~M. Stewart, ``Text as data: The promise and pitfalls of
  automatic content analysis methods for political texts,'' \emph{Political
  Analysis}, vol.~21, no.~3, pp. 267--297, 2013.

\bibitem{Han}
M.~K. Han~Jiawei and J.~Pei, \emph{Data Mining: Concepts and Techniques}, 2011.

\bibitem{voice}
P.~Melville, V.~Chenthamarakshan, R.~D. Lawrence, J.~Powell, and M.~Mugisha,
  ``Amplifying the voice of youth in africa via text analytics,'' in \emph{KDD
  '13 Proceedings of the 19th ACM SIGKDD international conference on Knowledge
  discovery and data mining}, 2013.

\bibitem{Cornell}
B.~Pang, L.~Lee, and S.~Vaithyanathan, ``Thumbs up? sentiment classification
  using machine learning techniques,'' in \emph{EMNLP '02 Proceedings of the
  ACL-02 conference on Empirical methods in natural language processing}, 2002.

\bibitem{agarwal:twitter}
A.~Agarwal, B.~X.~I. Vovsha, O.~Rambow, and R.~Passonneau, ``Sentiment analysis
  of twitter data,'' 2011.

\bibitem{ucla:image}
J.~Joo, W.~Li, F.~F. Steen, , and S.-C. Zhu, ``Visual persuasion: Inferring
  communicative intents of images,'' 2014.

\bibitem{Illonois}
B.~Liu, M.~Hu, and J.~Cheng, ``Opinion observer: Analyzing and comparing
  opinions on the web,'' in \emph{Proceedings of the 14th international
  conference on World Wide Web}, 2005.

\bibitem{Krasner}
S.~D. Krasner, \emph{Sovereignty: Organized Hypocrisy}.\hskip 1em plus 0.5em
  minus 0.4em\relax Princeton University Press, 1999.

\bibitem{Peeters:Negativity}
G.~Peeters and J.~Czapinski, ``Positive-negative asymmetry in evaluations: The
  distinction between affective and informational negativity effects,''
  \emph{European Review of Social Psychology}, vol.~1, pp. 33--60, 1990.

\bibitem{roy:bad}
R.~F. Baumeister, E.~Bratslavsky, K.~D. Vohs, and C.~Finkenauer, ``Bad is
  stronger than good,'' \emph{Review of General Psychology}, vol. Vol. 5., no.
  No. 4., pp. 323--370, 2001.

\bibitem{eubusiness}
\BIBentryALTinterwordspacing
B.~Hall and G.~Dyer. (2008, 11) Business fears over chinese-french rift.
  [Online]. Available:
  \url{http://www.ft.com/cms/s/0/107fbab4-bbf2-11dd-80e9-0000779fd18c.html}
\BIBentrySTDinterwordspacing

\bibitem{Huntington:Civilization}
S.~P. Huntington, \emph{The Clash of Civilizations and the Remaking of World
  Order}.\hskip 1em plus 0.5em minus 0.4em\relax Simon \& Schuster, 2011.

\bibitem{nye}
J.~S. Nye, ``American and chinese power after the financial crisis,''
  \emph{Washington Quarterly}, vol.~33, no.~4, pp. 143--153, 2010.

\bibitem{eastasia}
B.~J. Cohen and E.~M. Chiu, Eds., \emph{Power in a Changing World Economy:
  Lessons from East Asia}.\hskip 1em plus 0.5em minus 0.4em\relax Routledge,
  2013.

\bibitem{UK-US}
J.~Dumbrell, ``The us-uk special relationship: Taking the 21st-century
  temperature,'' \emph{The British Journal of Politics and International
  Relations}, vol.~11, pp. 64--78, 1 2009.

\bibitem{Alexander:Germany}
A.~G. Rahr, ``Germany and russia: A special relationship,'' \emph{The
  Washington Quarterly}, vol.~30, no.~2, pp. 137--145, 2007.

\bibitem{Koo:Japan}
M.~G. Koo, ``The senkaku/diaoyu dispute and sino-japanese political-economic
  relations: cold politics and hot economics?'' \emph{The Pacific Review},
  vol.~22, no.~2, pp. 205--232, 2009.

\bibitem{wsj:hongkong}
\BIBentryALTinterwordspacing
I.~Steger, F.~Law, and P.~Ho. (2014, 9) Pro-democracy protests shake hong kong.
  [Online]. Available:
  \url{http://www.wsj.com/articles/hong-kongs-pro-democracy-protesters-face-off-against-police-1411876944}
\BIBentrySTDinterwordspacing

\bibitem{world_bank}
\BIBentryALTinterwordspacing
S.~Jiang. (2014, 10) World bank says occupy protests fail to impact hong kong's
  business climate. [Online]. Available:
  \url{http://www.scmp.com/business/economy/article/1627626/world-bank-says-occupy-protests-fail-impact-hks-business-climate}
\BIBentrySTDinterwordspacing

\bibitem{abridgenottoofar}
\BIBentryALTinterwordspacing
(2015, 3) A bridge not far enough. The Economist. [Online]. Available:
  \url{http://www.economist.com/news/leaders/21646746-america-wrong-obstruct-chinas-asian-infrastructure-bank}
\BIBentrySTDinterwordspacing

\end{thebibliography}

\end{document}